\documentclass[a4paper]{jpconf}
	\usepackage{graphicx}
	\def\be{\begin{equation}}
	\def\ee{\end{equation}}
	\def\l{\left} 	\def\r{\right}
	\def\pa{\partial}
	\newcommand{\eq}[1]{(\ref{#1})}
\begin{document}

\title{The cool potential of gluons}

\author{Andr\'e Peshier$^\star$}
\address{Department of Physics, University of Cape Town, Rondebosch 7700, South Africa}
\ead{Andre.Peshier@uct.ac.za}

\author{Dino Giovannoni}
\address{Department of Physics and Electronics, Rhodes University, Grahamstown 6140, South Africa}
\ead{D.Giovannoni@ru.ac.za}

\begin{abstract}
We put forward the idea that the quark-gluon plasma might exist way below the usual confinement temperature $T_c$.
Our argument rests on the possibility that the plasma produced in heavy-ion collisions could reach a transient quasi-equilibrium with `over-occupied' gluon density, as advocated by Blaizot et al. 
Taking further into account that gluons acquire an effective mass by interaction effects, they can have a positive chemical potential and therefore behave similarly to non-relativistic bosons. Relevant properties of this dense state of interacting gluons, which we dub serried glue, can then be inferred on rather general grounds from Maxwell's relation.
\end{abstract}

\section{Introduction}
Experiments at RHIC and LHC have firmly established that the quark-gluon plasma produced in heavy ion collisions approaches local equilibrium rapidly, at a rate which allows for successful hydrodynamical descriptions of the parton phase despite its short lifetime of some fm/c.
It seems reasonable to consider that this approach to equilibrium is initially dominated by binary collisions of partons \cite{Blaizot:2011xf}.
If indeed gluon number changing processes cannot keep up with the fast kinetic equilibration, the quark-gluon plasma might exit sufficiently long in a {\em transient} quasi-equilibrium state that is characterized not only by local temperature, collective flow velocity and quark chemical potential(s), but also by a non-zero {\em gluon} chemical potential $\mu \not= 0$, quantifying that the gluon phase space density has not yet reached the asymptotic Bose distribution with $\mu_\infty = 0$.
As a consequence thereof, it has been argued that the high gluon occupation in the initial state, which can be described as a color glass condensate, may lead to the formation of a gluonic Bose-Einstein condensate in heavy ion collisions \cite{Blaizot:2013lga}.

Here we will put forward another noteworthy implication of this {\em transient} quasi-equilibrium phase with gluon overpopulation.
In a thermal medium massless partons acquire a dynamical mass $m \sim gT$, which is a key mechanism to understand thermodynamic properties of the strongly interacting quark-gluon plasma, see e.\,g.\ \cite{Peshier:1995ty}.
In the present context it implies that the chemical potential $\mu$, which must be smaller than the minimal energy of the dressed gluons $\omega(k) = (k^2+m^2)^{1/2}$, can be positive (opposed to $\mu \le 0$ assumed in \cite{Blaizot:2013lga}, and in general in perturbative QCD). This {\em dense and interacting} phase at $0 < \mu < m$ we dub {\bf serried glue}.

\begin{figure}[ht]
\begin{minipage}{8cm}
\caption{\small Schematic `phase diagram' of the gluon plasma in {\em transient} equilibrium with $\mu \not= 0$. Only with the dynamical gluon mass $m$ taken into account, the window for the serried glue phase opens between the under-occupied phase at $\mu<0$ and the Bose-Einstein condensate at $\mu_{_{\rm BEC}} = {\rm min}[\omega(k)] = m$. \label{PhaseDiagr schematic}}
\end{minipage}\hfill
\begin{minipage}{7cm}
\includegraphics[width=75mm]{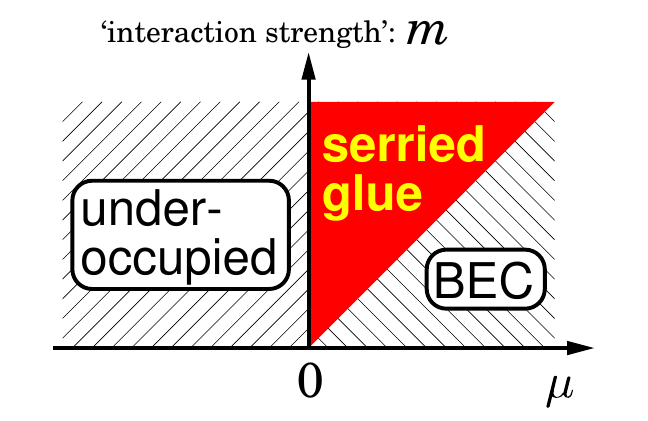}
\end{minipage} 
\end{figure}
\vspace{-5mm}

\section{QCD quasiparticles and Maxwell flow}
In order to study the thermodynamic properties of the serried glue phase, we generalize a QCD quasiparticle model \cite{Peshier:1995ty}: In the {\em fully} equilibrated system, i.\,e.\ at $\mu_\infty = 0$, dressed gluons can be approximated as quasiparticles with a temperature-dependent mass $m^2(T) = \textstyle\frac12\, G^2(T) T^2$ where 
\be
	G^2(T) = b / \ln\! \Big( (\lambda T + T_s)/T_c \Big)^2 
	\label{G2}
\ee
is an {\em effective} coupling, with $b = 16\pi^2/(11-\frac23\,n_{\! f})$ for $n_{\! f}$ flavors. 
Thermodynamic properties follow from the thermodynamic potential $-p(T)V$. For simplicity we focus here on the quenched limit, $n_{\! f} = 0$; then the quasiparticle pressure is that of an ideal gas of bosons with mass $m(T)$ and degeneracy $d_g = 16$, less a residual quasiparticle interaction contribution,
\be
	p(T) = p_{\rm id}(T; m(T)) - B(T) \, .
\ee
By thermodynamic consistency, $B(T)$ is related to (and can be obtained, up to an integration constant, from) the quasiparticle mass $m(T)$ by
\[
	\frac{dB(T)}{dT} = \l.\frac{\pa p_{\rm id}(T; m)}{\pa m}\r|_T \frac{dm(T)}{dT} \, .
\]
As consequence thereof, the entropy density takes a particularly simple form (without a residual interaction term like in the pressure), viz.
\be
	s(T) = \pa p(T)/\pa T = s_{\rm id}(T; m(T)) \, .
	\label{s}
\ee
After fixing the parameters $\lambda$ and $T_s$ by matching to lattice QCD results \cite{Borsanyi:2012ve}, the scheme is fully specified and will provide the `initial' condition for a flow equation derived now.

\medskip
For $\mu \not= 0$, both the quasiparticle mass and the pressure will depend on the chemical potential. 
Similar to the entropy density (\eq{s} is now complemented by a $\mu$-dependence), we find the gluon particle density equal to the density of an ideal boson gas with ${T,\mu}$-dependent mass,
\be
	n(T,\mu) = \pa p(T,\mu)/\pa\mu = n_{\rm id}(T, \mu; m(T,\mu)) \, .
	\label{n}
\ee
Obviously, as derivatives of $p(T,\mu)$, the entropy and particle density must satisfy Maxwell's relation $\pa s/\pa \mu = \pa n/\pa T$, which translates into a condition for the quasigluon mass $m(T, \mu)$,
\be
	\frac{\pa m}{\pa \mu} + A\Big( T, \mu; m(T, \mu) \Big)\, \frac{\pa m}{\pa T} = 0 \, .
	\label{flow eq}
\ee
This flow equation is similar to a relation derived in \cite{Peshier:1999ww} for the effective coupling $G$ in the quark-gluon plasma at finite {\em quark} chemical potential $\mu_Q$, which led to an early successful prediction of the curvature of the critical line to the hadronic phase in the $T \mu_Q$ phase diagram.

\begin{figure}[ht]
\begin{minipage}{88mm}
\caption{\small Illustrating the Maxwell flow \eq{flow eq}: the end points of the characteristics specify the boundary to the Bose-Einstein condensate (dotted green line). The charac\-teristic line $\mu^c(T, T_c)$ from the `critical point' $(T_c,0)$ can be roughly related to the confinement phase boundary.
Note that the characteristics from a vicinity of $T_c$ intersect (see dash-dotted orange line) since the solution of \eq{flow eq} is locally non-unique: $m(T,\mu)$ develops a fold (as explained in the text, this has no physics relevance).
\label{CharLines}}
\end{minipage}\hfill
\begin{minipage}{7cm}
\includegraphics[width=75mm]{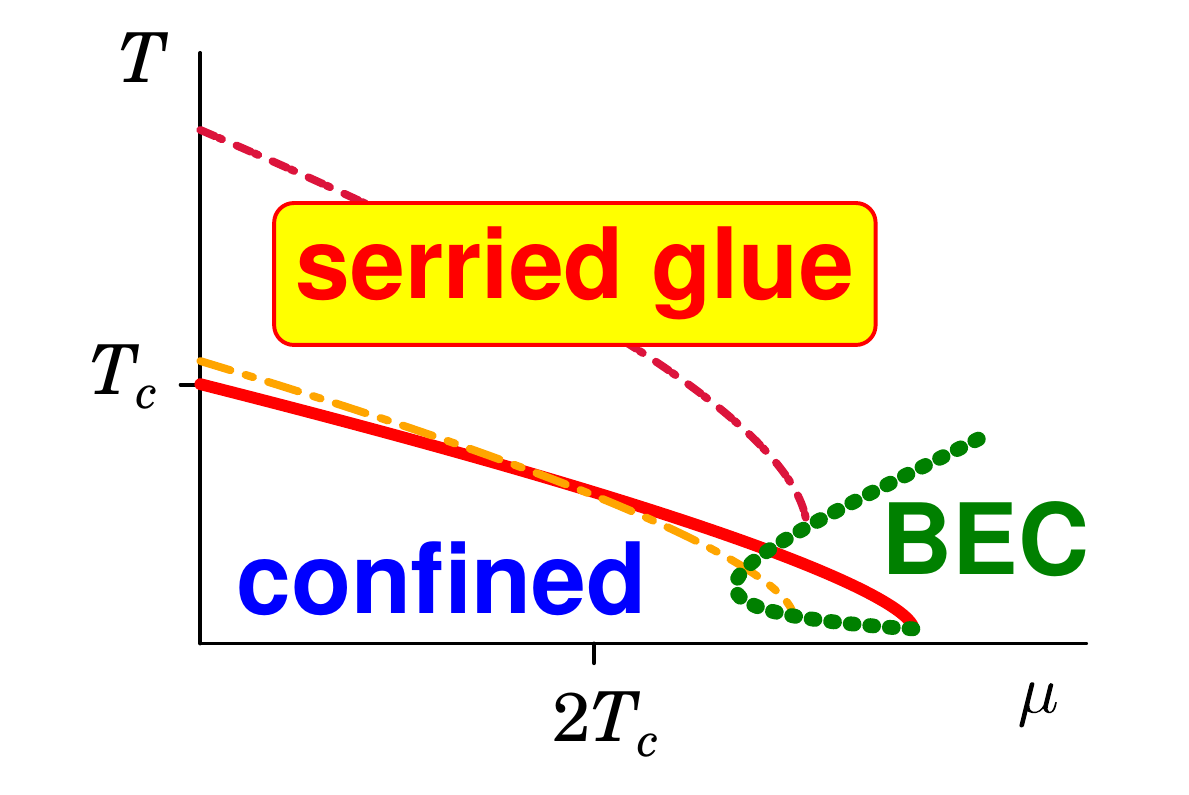}
\end{minipage} 
\end{figure}

We only mention (without detailing it here) that the expression $A$ in \eq{flow eq} is a non-linear function of $T, \mu, m$ involving integrals of the gluon distribution $f(k) = (\exp[(\omega(k)-\mu)/T] - 1)^{-1}$. Note that nonetheless \eq{flow eq} is a {\em linear} partial differential equation which can be readily solved by the {\sl method of characteristics} (with the `initial' condition $m(T,0)$ obtained from fitting the usual $\mu=0$ lattice QCD results, as announced).
We then notice immediately that with only derivatives of $m$ in the flow equation \eq{flow eq}, i.\,e.\ no summand term $\tilde A(T, \mu; m )$, the mass $m(T,\mu)$ is {\em constant} along the characteristics. Therefore, the characteristic line  $\mu^c(T, T_i)$ from a given `initial' point $(T_i,0)$ into the $\mu > 0$ region is limited by $\mu \le \mu^{\rm max}(T_i)=m(T_i,0)$. This upper bound  $\mu^{\rm max}$ is in fact reached (with vanishing slope $\pa\mu^c(T, T_i)/\pa T$) and earmarks the onset of Bose-Einstein condensation.
The corresponding temperatures at the end points of the characteristic lines, which are needed to fully specify the condensate phase boundary $\mu_{_{\rm BEC}}^\star(T)$, can be be calculated analytically in the weak-coupling limit or numerically from the flow equation \eq{flow eq}. As a qualitative feature we point out that the inverse relation $T_{_{\rm BEC}}^\star(\mu)$ has two branches, see Fig.~\ref{CharLines} -- an obvious result of the continuous mapping by the flow \eq{flow eq} of the `initial' condition $m(T,0) \propto G(T) T$ which is {\em non-monotonous} (being the product of a decreasing and an increasing function).
\vspace{-5mm}

\section{Results and conclusion}
Although Bose-Einstein condensation in heavy-ion collisions is certainly interesting, we focus here on another aspect. 
Sufficient for our argument, and easy to show analytically from \eq{flow eq}, is that the slope of the characteristics is negative. Following \cite{Peshier:1999ww} and relating the characteristic line $\mu^c(T, T_c)$ from the `critical point' $(T_c,0)$ to the phase boundary with the confined phase, we find that the confinement temperature $T_c^\star(\mu)$ decreases markedly with $\mu$ in the serried glue phase, see full red line in Fig.~\ref{CharLines}.

\begin{figure}[ht]
\centerline{\includegraphics[width=98mm]{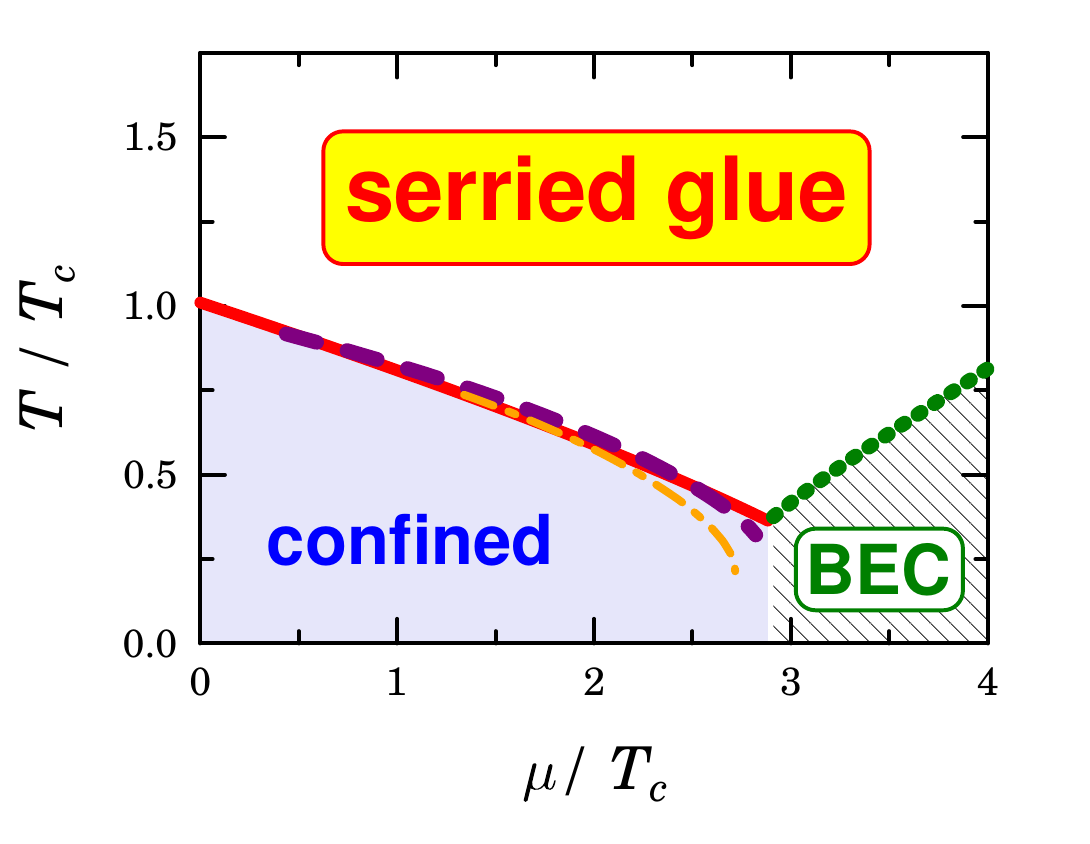}} \vskip -5mm
\caption{\small Proposed phase diagram of (the gluon sector) of QCD matter. As discussed in the text, the system will undergo a transition to the confined phase `before' it reaches on the descending characteristics $p(T,\mu) = 0$ (indicated by the dashed purple line), hence the multi-valued solution of \eq{flow eq} in a region below the dash-dotted orange line is irrelevant.
\label{PhaseDiagram}}
\end{figure}

In this context we remark that for large effective coupling the solution of \eq{flow eq} turns out to be multi-valued in a small region, indicated by characteristics from a small vicinity of $T_c$ intersecting each other, cf.\ Fig.~\ref{CharLines}. However, this circumstance has no physical relevance since the pressure, when calculated `along' these descending characteristics, becomes negative `before' the solution of \eq{flow eq} becomes non-unique. This is illustrated in Fig.~\ref{PhaseDiagram} where $p(T,\mu) = 0$ is indicated by the dashed purple line -- a `conservative bound' for the confinement transition occurring in fact at non-zero pressure (i.\,e.\ somewhat above the dashed line), which thus excludes seemingly ambiguous solutions of the flow equation.We recall a corresponding observation in case of the quark chemical potential \cite{Peshier:1999ww}.

Of interest for phenomenology are quantitative estimates of the magnitude of the expected effects. Keeping in mind that for simplicity we are discussing here the quenched limit where $T_c \simeq 0.26\,$GeV, we may expect a pure gluon plasma in the serried phase to exist down to temperatures of some $0.1\,$GeV, if the gluon chemical potential reaches $\mu^\star \sim 3T_c$ where Bose-Einstein condensation would set in.
We remark that such a {\sl `massive'} effect\footnote{
	Even though it might be slightly misleading, we are tempted to call it gluon {\em `super-cooling'}, in particular because it is also a transient phenomenon like the usual super-cooling.}
is not a surprise: massive bosons with chemical potential close to $\mu^{\rm max} = m$ behave like non-relativistic particles, $\omega(k) - \mu \to (k^2+m^2)^{1/2} - m = k^2/2m + \ldots$ Thus the `Bose divergence', $(e^x-1)^{-1} \approx x^{-1}$ for small $x$, is enhanced compared to perturbatively massless gluons ($\mu^{\rm max} \to 0$), which can over-compensate the suppression effects due to larger kinetic energy with a dynamical mass.

\smallskip
In conclusion, we propose a new phase `serried glue' in the QCD phase diagram. 
While we discussed here only the quenched limit, we expect also for the physical case the possibility of reaching  temperatures significantly below the commonly assumed lowest value $T_c \approx 0.16\,$GeV, catalyzed by interacting (i.\,e.\ dressed) gluons in over-occupied states \cite{DG+AP}. We consider this idea as reasonably robust insofar as it relies on the indubitable emergence of dynamical mass scales in the medium and the fundamental Maxwell relation (which governs not only the quasiparticle model used here, but also more sophisticated resummation approaches like \cite{Peshier:2000hx}).
The only remaining assumption, the existence of a sufficiently long-lived transient equilibrium with approximately conserved gluon number, remains to be substantiated. If confirmed, the implications of a `super-cool serried glue' phase would be far-reaching for hydrodynamic simulations and thus for heavy-ion physics in general.

\end{document}